\documentclass[11pt]{article}
\pdfoutput=1
\usepackage{geometry}                
\usepackage{graphicx}
\usepackage{amsmath}
\usepackage{amssymb}
\usepackage{authblk}

\usepackage[scr=boondoxo,scrscaled=1.05]{mathalfa}

\usepackage{epstopdf}
\usepackage[backend=bibtex,style=numeric,sorting=none]{biblatex}
\renewbibmacro{in:}{}
\DeclareFieldFormat[article]{volume}{\mkbibbold{#1}}
\ifdefined\blackandwhite
\usepackage[pdftex]{hyperref} 
\else
\usepackage[colorlinks,pdftex]{hyperref} 
\hypersetup{citecolor = blue, linkcolor = red} 
\fi

\newcommand{\bra}[1]{\left\langle #1 \right|}
\newcommand{\ket}[1]{\left|#1\right\rangle}

\newcommand{\brakets}[3]{\left\langle#1 | #2 |  #3\right\rangle}

\title{Classification with  Quantum Neural Networks\\on Near Term Processors}
\author{Edward Farhi$^{1,2}$ and Hartmut Neven$^{1}$}
\affil{$^{1}${Google Inc.\\Venice, CA 90291}\\ \vspace{1ex} $^{2}${Center for Theoretical Physics\\ Massachusetts Institute of Technology\\Cambridge, MA 02139}}
\date{}    

\begin{document}

\maketitle

\begin{abstract}
We introduce a quantum neural network, QNN, that can represent labeled data, classical or quantum, and be trained by supervised learning.  The quantum circuit consists of a sequence of parameter dependent unitary transformations which acts on an input quantum state.  For binary classification a single Pauli operator is measured on a designated readout qubit.  The measured output is the quantum neural network's predictor of the binary label of the input state.  First we look at classifying classical data sets which consist of n-bit strings with binary labels.  The input quantum state is an n-bit computational basis state corresponding to a sample string. We show how to design a circuit made from two qubit unitaries that can correctly represent the label of any Boolean function of n bits.  For certain label functions the circuit is exponentially long.  We introduce parameter dependent unitaries that can be adapted by supervised learning of labeled data.  We study an example of real world data consisting of downsampled images of handwritten digits each of which has been labeled as one of two distinct digits.  We show through classical simulation that parameters can be found that allow the QNN to learn to correctly distinguish the two data sets.  We then discuss presenting the data as quantum superpositions of computational basis states corresponding to different label values. Here we show through simulation that learning is possible.  We consider using our QNN to learn the label of a general quantum state. By example we show that this can be done. Our work is exploratory and relies on the classical simulation of small quantum systems. The QNN proposed here was designed with near-term quantum processors in mind. Therefore it will be possible to run this QNN on a near term gate model quantum computer where its power can be explored beyond what can be explored with simulation. 
\end{abstract}

\section{Introduction and Setup}

Artificial intelligence in the form of machine learning has made great strides towards getting classical computers to classify data \cite{LeCun}\cite{Ian}.  Here we imagine a large data set consisting of strings where each string comes with a binary label.  For simplicity we imagine that there is no label noise so that we can be confident that the label attached to each string is correct.  We are given a training set which is a set of $S$ samples of strings with their labels. The goal is to use this information to be able to correctly predict the labels of unseen examples.  Clearly this can only be done if the label function has underlying structure.  If the label function is random we may be able to learn (or fit with $S$ parameters) the labels from the training set but we will not be able to say anything about the label of a previously unseen example.  Now imagine a real world example where the data set consists of pixilated images each of which has been correctly labeled to say if there is a dog or a cat in the image.  In this case classical neural networks can learn to correctly classify new images as dog or cat.  We will not review how this is done in the classical setting but rather turn immediately to a quantum neural network capable of learning to classify data.  We continue to use the word ``neural" to describe our network since the term has been adopted by the machine learning community recognizing that the connection to neuroscience is now only historical. Other approaches to harnessing quantum resources in machine learning are reviewed here \cite{QML1}\cite{QML2}.

To be concrete, imagine that the data set consists of strings $z= z_1 z_2 \ldots z_n$ where each $z_i$ is a bit taking the value $+1$ or $-1$ and a binary label $l(z)$ chosen as $+1$ or $-1$. For simplicity imagine that the data set consists of all $2^n$ strings.  We have a quantum processor that acts on $n+1$ qubits and we ignore the possible need for ancilla qubits.  The last qubit will serve as a readout. The quantum processor implements unitary transformations on input states.  The unitaries that we have come from some toolbox of unitaries, perhaps determined by experimental considerations \cite{2017arXiv170306199F}.  So we have a set of basic unitaries
\begin{equation}
\left\{U_a (\theta)\right\} 
\label{farhar1}
\end{equation}                 
each of which acts on a subset of the qubits and depends on a continuous parameter $\theta$, where for simplicity we have only one control parameter per unitary. Now we pick a set of $L$ of these and make the unitary
\begin{equation}
U(\vec{\,\theta}\, ) = U_L (\theta_L) \, U_{L-1} (\theta_{L-1}) \ldots   U_1 (\theta_1)
\label{farhar2}
\end{equation}
which depends on the $L$ parameters $\vec{\theta} = \theta_L, \theta_{L-1},\dots\theta_1$.  For each $z$ we construct the computational basis state 
\begin{equation}
\ket{z, 1} = \ket{z_1\, z_2 \ldots z_n ,1}
\label{farhar3}
\end{equation}
where the readout bit has been set to 1.   Acting with the unitary on the input state gives the state
\begin{equation}
U(\vec{\,\theta}\, ) \ket{z, 1}.
\label{farhar4}
\end{equation}
On the readout qubit we then measure a Pauli operator, say $\sigma_y$, which we call $Y_{n+1}$.  This gives a $+1$ or a $-1$.  Our goal is to make the measurement outcome correspond to the correct label of the input string, that is, $l(z)$. Typically the measurement outcome is not certain. Our predicted label value is the real number between $-1$ and $1$,
\begin{equation}
\bra{z , 1} U^\dagger (\vec{\,\theta}\, ) Y_{n+1} U(\vec{\,\theta}\, ) \ket{z , 1} 
\label{farhar4b}
\end{equation}
which is the average of the observed outcomes if $Y_{n+1}$ is measured in multiple copies of \eqref{farhar4}.

\begin{figure}
  \centering
  {\includegraphics[width=\textwidth]{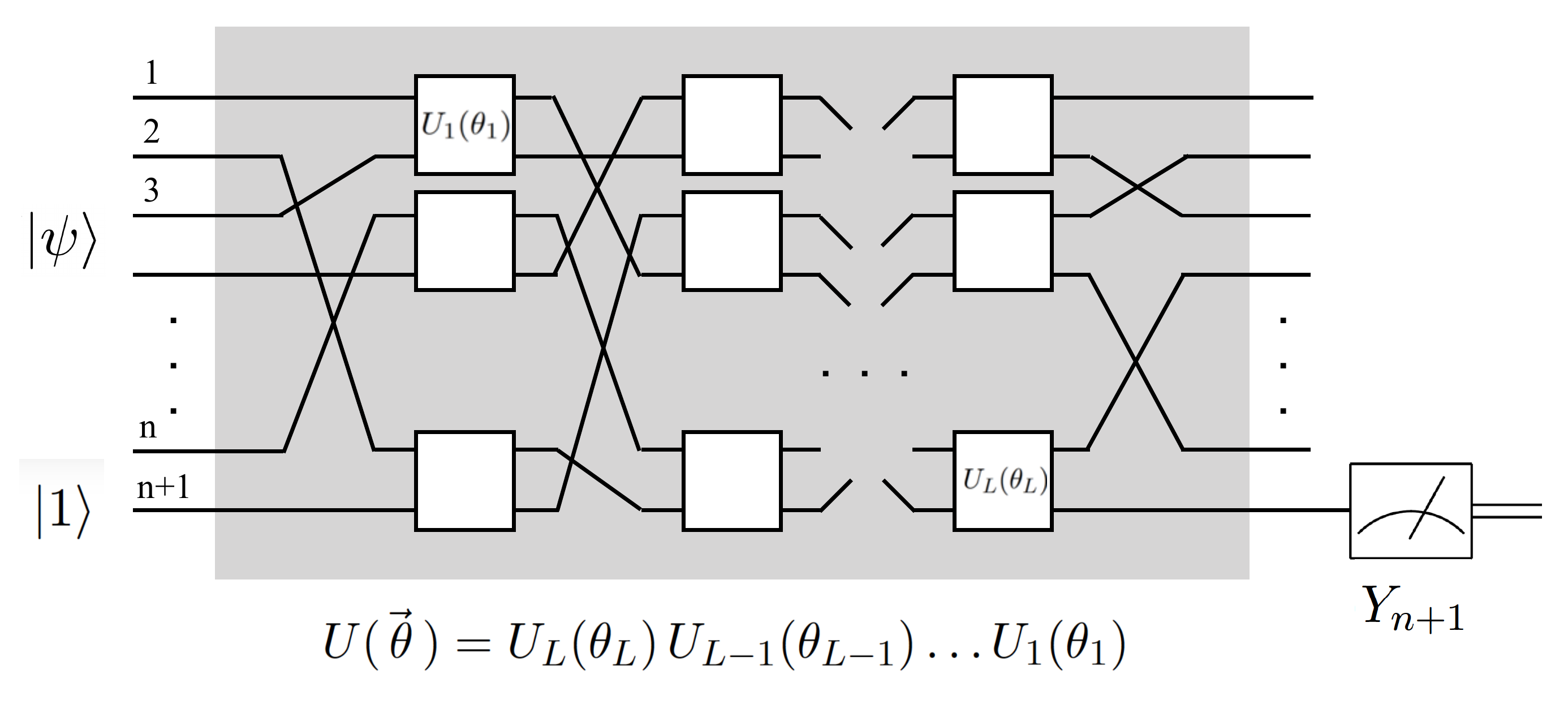}}
  \caption{Schematic of the proposed quantum neural network on a quantum processor. The input state $\ket{\psi, 1}$ is prepared and then transformed via a sequence of few qubit unitaries $U_i(\theta_i)$ that depend on parameters $\theta_i$. These get adjusted during learning such that the measurement of $Y_{n+1}$ on the readout qubit tends to produce the desired label for $\ket{\psi}$.}
\label{Figure1}
\end{figure}

Our goal is to find parameters $\vec{\theta}$ so that the predicted label is near the true label. We will address the question of whether such parameters even exist (representation) as well as the question of whether such optimal parameters can then be efficiently found (learning).  For a given circuit, that is, a choice of $L$ unitaries, and a set of parameters $\vec{\theta}$, and an input string $z$, consider the sample loss
\begin{equation}
\text{loss} (\vec{\theta}, z) = 1 - l(z) \bra{z , 1} U^\dagger (\vec{\,\theta}\, ) Y_{n+1} U(\vec{\,\theta}\, ) \ket{z , 1}.
\label{farhar5}
\end{equation}
Note that the sample loss we use is linear in the margin (the product of the label and the predicted label value) and its minimum is at 0 (not minus infinity) because the predicted label value is automatically bounded to be between -1 and 1.   Suppose that the quantum neural  network is working perfectly, so that for each input $z$, the measurement always gives the correct label. This would mean that parameters $\vec{\theta}$ exist and have been found such that the sample loss is 0 for all inputs $z$. 

Given a training set of $S$ strings with their labels we now describe how to use the quantum processor to find the parameters $\vec{\theta}$ that achieve the learning task, that is, we describe supervised learning on a quantum neural network.  For now we are assuming that our circuit is rich enough that there exist parameters that allow us to  represent the label.  Start with say random parameters $\vec{\theta}$ or perhaps an inspired choice.  Pick a string $z^1$ from the training set.  Use the quantum processor to construct
\begin{equation}
U (\vec{\,\theta}\, ) \ket{z^1, 1}
\label{farhar6}
\end{equation}
and measure $\sigma_y$ on the last qubit.  Do this enough times to get a good estimate of the expected value of $Y_{n+1}$ and then compute loss$(\vec{\theta}, z^1)$ via equation \eqref{farhar5}.  Now we want to make a small change in the parameters $\vec{\theta}$ to reduce the loss on training example $z^1$.  We might do this by randomly sampling from nearby $\vec{\theta}$'s.  Or we could compute the gradient with respect to $\vec{\theta}$ of loss$(\vec{\theta}, z^1)$ and then take a small step in the direction that reduces the loss. (More on how to get the gradient later.)  This gives us  new parameters $\vec{\theta}^{\,1}$.  We now take a new training example $z^2$ and with quantum measurements estimate loss$(\vec{\theta}^{\,1}, z^2)$.  Now change $\vec{\theta}^{\,1}$ by a little bit to slightly reduce this loss.  Call the new parameters $\vec{\theta}^{\,2}$.  We can continue in this fashion on say the whole set of training samples generating a sequence $\vec{\theta}^{\,1}, \vec{\theta}^{\,2}\ldots\vec{\theta}^{\,S}$.  If the learning has been successful we would find that with the parameters $\vec{\theta}^S$, the operator $U(\vec{\theta}^S)$ acting on the state $\ket{z , 1}$  will produce a state which when the output qubit is measured will give the correct label $l(z)$.  If $z$ is from the training set we could claim that we have fit the training data.  If $z$ is outside the training set, say from a specified test set, we would say that the learning has generalized to unseen examples.

What we have just described is an implementation of what in classical machine learning is called ``stochastic gradient descent."  The stochasticity comes from the fact that the training examples are drawn randomly  from the training set. When learning is successful, after enough training examples are processed, the parameters settle into a place where labels can be correctly predicted.  There may be many values of the parameters that result in success and for this reason it may be that even if the number of parameters is very large, starting from a random point can lead to a good solution.  In traditional machine learning with a neural network, the parameters (called weights) appear as entries in matrices that act linearly on internal vectors.  The components of these vectors are acted on non-linearly before the vector is multiplied by other weight dependent matrices.  Part of the art of building a successful machine learning implementation is the introduction of the right non-linearity.  In our setup each unitary acts on the output of the previous unitary and no non-linearities are explicitly introduced.  What we specify is the set of parameterized unitaries and the operator to be measured after the quantum evolution.  Imagine that the individual unitaries in the set \eqref{farhar1} are all of the form 
\begin{equation}
\exp\, (i\, \theta\, \Sigma)
\label{farhar7}
\end{equation}
where $\Sigma$ is a generalized Pauli acting on a few qubits, that is, $\Sigma$ is a tensor product of operators from the set $\left\{\sigma_x, \sigma_y, \sigma_z\right\}$ acting on a few qubits.  The derivative with respect to $\theta$ gives an operator whose norm is bounded by 1. Therefore the gradient of the loss function with respect to $\vec{\theta}$ is bounded by $L$, the number of parameters. This means that the gradient cannot blow up and in this way we avoid a well known problem that can occur when computing gradients in classical neural networks. 
Researchers in classical machine learning recently started to investigate the advantage of using unitary transformations to control gradient blow up \cite{2015arXiv151106464A}\cite{2017arXiv171011160D}\cite{2017arXiv170509792T}\cite{2016arXiv160704903H}. Note that in our case this advantage comes for free.
\section{Representation}
Before discussing learning we want to establish that our quantum neural network is capable of expressing any two valued label function, although as we will see at a possibly high cost in circuit depth (see also \cite{2014NJPh...16j3014Y}).  There are $2^n$, $n$-bit strings and accordingly there are $2^{(2^n)}$ possible label functions $l(z)$.  Given a label function consider the operator whose action is defined on computational basis states as
\begin{equation}
U_l \ket{z, z_{n+1}} = \exp \big(i  \tfrac{\pi}{4}\, l(z) X_{n+1}\big) \ket{z, z_{n+1}} . 
\label{farhar8}
\end{equation}
In other words it acts by rotating the output qubit about its x-axis by $\frac{\pi}{4}$ times the label of the string $z$.  Correspondingly, 
\begin{equation}
U_l^\dagger \, Y_{n+1} U_l  =  \cos \big(\tfrac{\pi}{2}\ l(Z) \big)  Y_{n+1}  +  \sin \big(\tfrac{\pi}{2}\ l(Z) \big)  Z_{n+1}
\label{farhar9}
\end{equation}
where in this formula $l(Z)$ is interpreted as an operator diagonal in the computational basis.   Note that since $l(z)$ can only be $+1$ or $-1$ we have that
\begin{equation}
\bra{z , 1} U_l^\dagger\,Y_{n+1} U_l  \ket{z, 1} = l(z)  .
\label{farhar10}
\end{equation}
This shows that at least at some abstract level we have a way of representing any label function with a quantum circuit.  

We now show how to write $U_l$ as a product of two qubit unitaries.  For this discussion it is convenient to switch to Boolean variables $b_i=\frac{1}{2} (1 - z_i)$ and think of our label function $l$ as $1 - 2 b$ where $b$ is 0, 1 valued.   Now we can use the Reed-Muller representation of any Boolean function in terms of the bits $b_1$ through $b_n$:  
\begin{equation}
b = a_0 \oplus (a_1\, b_1 \oplus a_2\, b_2 \oplus\ldots a_n\, b_n) \oplus (a_{12}\, b_1\, b_2 \oplus a_{13}\, b_1\, b_3 +\ldots) \oplus \ldots \oplus a_{123}\ldots b_1\, b_2\ldots b_n.
\label{farhar11}
\end{equation}
The addition is mod2 and the coefficients $a$ are all 0 or 1.  Note that there are $2^n$ coefficients and since they are each 0 or 1 we see that there are indeed $2^{(2^n)}$ Boolean functions being represented.  The formula can be exponentially long.   Now we can write the label dependent unitary  $U_l$ in \eqref{farhar8} as 
\begin{equation}
 U_l   = \exp( i\,  \tfrac{\pi}{4}\,  X_{n+1}) \exp ( -i\, \tfrac{\pi}{2}\, B\, X_{n+1})  
 \label{farhar12}
\end{equation}
where $B$ is the operator, diagonal in the computational basis, corresponding to $b$.  Each term in $B$ in \eqref{farhar12} is multiplied by $X_{n+1}$ and so each term commutes with the others.  Each non-vanishing term in the Reed-Muller formula gives rise in $U_l$ to a controlled bit flip on the output qubit.  To see this consider say the three bit term involving  bits 2, 7 and 9.  This corresponds to the operator
\begin{equation}
\exp (-i\,  \tfrac{\pi}{2}\, B_2 \, B_7 \, B_9 \, X_{n+1})
 \label{farhar13}
\end{equation}
which, acting on a computational basis state on the first $n$ qubits, is the identity unless $b_2=b_7=b_9=1$ in which case it is $-i \, X_{n+1}$.  We know from early work \cite{Elementary} that any controlled one qubit unitary acting on qubit $n+1$ where the control is on the first $n$ bits can be written as a product of $n^2$ two qubit unitaries.   So any label function expressed in terms of the Reed-Muller formula with say RM terms can be written as a product of RM commuting $n+1$ qubit operators and each of these can be written as $n^2$ two qubit unitaries. 

Our quantum representation result is analogous to the classical representation theorem\cite{UniversalRep1}\cite{UniversalRep2}.  This states that any Boolean label function can be represented on a depth three neural network with the inner layer having size $2^n$.  Of course such gigantic matrices cannot be represented on a conventional computer. In our case we naturally work in a Hilbert space of exponential dimension but we may need exponential circuit depth to express certain functions. The question of which functions can be compactly represented on a quantum circuit whereas they cannot be on a classical network is an open area of investigation. To this end we now explore some examples.
\subsection{Representing Subset Parity and Subset Majority}
Consider the label function which is the parity of a subset of the bits.  Call the subset $\mathscr{S}$ and let $a_j= 1$ if $j$ is in the subset and $a_j = 0$ if $j$ is not in the subset.  The  Reed-Muller formula for the subset parity label is
\begin{equation}
P_{\mathscr{S}} (z) = \sum\limits_j \, \oplus \,  a_j\, b_j
 \label{farhar14}
\end{equation}
which  is just the linear part of \eqref{farhar11}, where again the addition is mod2.  This gives rise to the unitary that implements subset parity:
\begin{equation}
U_{P_{\mathscr{S}}} = \exp \big( i\, \tfrac{\pi}{4} \, X_{n+1} \big) \exp \Big( -i \,\tfrac{\pi}{2} \, \sum\limits_j a_j\, B_j X_{n+1} \Big)
 \label{farhar15}
\end{equation}
Note that in the exponent the addition is automatically mod2 because of the $\frac{\pi}{2}$ and the properties of $X_{n+1}$. The  circuit consists of (at most) $n$ commuting two qubit operators with the readout qubit  is in all of the two qubit gates.  Classically to represent  subset parity on a standard neural network requires three layers. 

Now consider the label function which is subset majority.  The label is 1 if the majority of the bits in the subset are 1 and the label is $-1$ otherwise.   It is easiest to represent subset majority using the $z$ variables.  Then the subset majority label can be written as
\begin{equation}
M_{\mathscr{S}} (z)  = \text{sign} \Big(\sum\limits_j a_j\, z_j\Big)
 \label{farhar16}
\end{equation}
where we assume that the size of the subset is odd to avoid an ambiguity that occurs in the even case if the sum is 0. Although this is a compact way of writing subset majority it is not in the Reed Muller form.  We can write subset majority in the form \eqref{farhar11} but \eqref{farhar16} is more convenient for our current discussion. 

Now consider the unitary 
\begin{equation}
U_{M_{\mathscr{S}}} = \exp \Big(i\, \tfrac{\beta}{2}\ \sum\limits_j a_j \, Z_j  X_{n+1} \Big)
 \label{farhar17}
\end{equation}
where we will specify $\beta$ momentarily. Conjugating $Y_{n+1}$ gives
\begin{equation}
U_{M_{\mathscr{S}}}^\dagger Y_{n+1} U_{M_{\mathscr{S}}}  = \cos \Big(\beta  \sum\limits_j a_j\, Z_j \Big)  Y_{n+1}  +  \sin\Big(\beta \sum\limits_j a_j\, Z_j \Big)  Z_{n+1}
 \label{farhar18}
\end{equation}
so we have that
\begin{equation}
\bra{z , 1} U_{M_{\mathscr{S}}}^\dagger Y_{n+1} U_{M_{\mathscr{S}}} \ket{z, 1}  = \sin\Big(\beta \sum\limits_j a_j\, Z_j\Big).
 \label{farhar19}
\end{equation}
The biggest that  $\sum\limits_j a_j z_j$ can be is $n$ so if we set $\beta$ equal to say $\frac{.9\pi}{n}$ we have that
\begin{equation}
\text{sign} \bigg(\sin \Big(\frac{.9\pi}{n} \sum_j a_j\, z_j \Big ) \bigg) = M_{\mathscr{S}} (z)  .
 \label{farhar20}
\end{equation}
This means that if we make repeated measurements of $Y_{n+1}$ and round the expected value up to 1 or down to $-1$ we can obtain perfect categorical error although the individual sample loss values are not 1 or $-1$.  In classical machine learning subset majority is an easy label to express because with one layer the labels on the whole data set can be separated with a single hyperplane.

\section{Learning}

In the introduction and setup section we discussed how with each new training example we need to modify the $\vec{\theta}$'s so that the sample loss is decreased.  We will now be explicit about two strategies to accomplish this although other strategies may do better. With the parameters $\vec{\theta}$ and a given training example $z$ we first estimate the sample loss given by \eqref{farhar5}. To do this we make repeated measurements of $Y_{n+1}$ in the state \eqref{farhar4}. To achieve, with probability greater than 99\%, an estimate of the sample loss that is within $\delta$ of the true sample loss we need to make at least $2/\delta^2$ measurements. 

Once the sample loss is well estimated we want to calculate the gradient of the sample loss with respect to $\vec{\theta}$.  A straightforward way to proceed is to vary the components of $\vec{\theta}$ one at a time.  With each changed component we need to recalculate loss$(\vec{\theta^\prime}, z)$ where $\vec{\theta^\prime}$ differs from $\vec{\theta}$ by a small amount in one component. Recall that one can get a second order accurate estimate of the derivative of a function by taking the symmetric difference,
\begin{equation}
 				\frac{df}{dx}(x) = \big(f(x+\epsilon)-f(x-\epsilon)\big)/(2\epsilon) + O(\epsilon^2).
\label{farhar20b}
\end{equation}
To achieve this you need to know that your error in the estimate of $f$ at each $x$ is no worse than $O(\epsilon^3)$.  To estimate $\text{loss} (\vec{\theta}, z )$ to order $\epsilon^3$ we need of order $1/\epsilon^6$ measurements. So, for instance, using the symmetric difference we can get each component of the gradient accurate to order $\eta$ by making of order $1/\eta^3$ measurements. This needs to be repeated $L$ times to get the full gradient.

There is an alternative strategy for computing each component of the gradient \cite{2017arXiv170102691R} which works when the individual unitaries are all of the form \eqref{farhar7}.  Consider the derivative of the sample loss \eqref{farhar5} with respect to $\theta_k$ which is associated with the unitary $U_k (\theta_k)$  which has the generalized Pauli operator $\Sigma_k$. Now
\begin{equation}
  \frac{d \text{loss}(\vec{\theta}, z)}{d \theta_k} = 2 \operatorname{Im}\big( \bra{z , 1} U_{1}^\dagger... U_{L}^\dagger Y_{n+1} U_L ...U_{k+1} \Sigma_k U_k ... U_{1} \ket{z, 1}\big)
\label{farhar20c}
\end{equation}
Note that $Y_{n+1}$ and $\Sigma_k$ are both unitary operators.  Define the unitary operator 
\begin{equation}
   \mathcal{U}(\vec{\theta})=  U_{1}^\dagger... U_{L}^\dagger Y_{n+1} U_L ...U_{k+1} \Sigma_k U_k ... U_{1}  
   \label{farhar20d}
\end{equation}
so we reexpress \eqref{farhar5} as 
\begin{equation}
   \frac{d \text{loss}(\vec{\theta}, z)}{d \theta_k} = 2 \operatorname{Im}\big( \bra{z , 1} \mathcal{U} \ket{z, 1}\big).
   \label{farhar20e}
\end{equation}
$\mathcal{U}(\vec{\theta})$ can be viewed as a quantum circuit composed of $2L+2$ unitaries each of which depends on only a few qubits. We can use our quantum device to let $\mathcal{U}(\vec{\theta})$ act on $\ket{z, 1}$. Using an auxiliary qubit we can measure the right hand side of \eqref{farhar20e}.  To see how this is done start with
\begin{equation}
				\ket{z, 1}  \frac{1}{\sqrt{2}}  \big( \ket{0} + \ket{1} \big)
\end{equation}
and act with $i \mathcal{U}(\vec{\theta})$ conditioned on the auxiliary qubit being 1.  This produces

\begin{equation}
				\frac{1}{\sqrt{2}} \big( \ket{z, 1} \ket{0} + i \mathcal{U}(\vec{\theta}) \ket{z , 1}  \ket{1}\big)
\end{equation}
Performing a Hadamard on the auxiliary qubit gives
\begin{equation}
				\frac{1}{2}\big( \ket{z , 1}  + i \mathcal{U}(\vec{\theta}) \ket{z , 1}  \ket{0} \big)  + \frac{1}{2}\big( \ket{z , 1}  - i \mathcal{U}(\vec{\theta}) \ket{z , 1} \ket{1}  \big) .
\end{equation}
Now measure the auxiliary qubit.  The probability to get $0$ is
\begin{equation}
					\frac{1}{2} - \frac{1}{2} \operatorname{Im}\big( \bra{z, 1} \mathcal{U}(\vec{\theta}) \ket{z, 1}  \big)
\end{equation}
so by making repeated measurements we can get a good estimate of the imaginary part which turns into an estimate of the $k$\textsc{\char13}th component of the gradient.  This method avoids the numerical accuracy issue that comes with approximating the gradient as outlined in the previous paragraph.  The cost is that we need to add an auxiliary qubit and run a circuit whose depth is $2L+2$.

Given an accurate estimate of the gradient we need a strategy for how to update $\vec{\theta}$.  Let $\vec{g}$ be the gradient of loss$(\vec{\theta}, z)$ with respect to $\vec{\theta}$.  Now we want to change $\vec{\theta}$ in the direction of $\vec{g}$.   To lowest order in $\gamma$ we have that
\begin{equation}
\text{loss} ( \vec{\theta} + \gamma \vec{g} , z ) = \text{loss} (\vec{\theta}, z) + \gamma \vec{g}^2  + \mathcal{O} (\gamma^2) .
 \label{farhar21}
\end{equation}
We want to move the loss to its minimum at 0 so the first thought is to make
\begin{equation}
\gamma = - \ \frac{\text{loss} (\vec{\theta}, z )}{\vec{g\,}^2} \  .
\label{farhar22}
\end{equation}
Doing this might drive the loss to near 0 for the current training example but in doing so it might have the undesirable effect of making the loss for other examples much worse.  The usual machine learning technique is to introduce a learning rate $r$ which is small and then set
\begin{equation}
\vec{\theta} \to  \vec{\theta} - r \bigg(\frac{\text{loss} (\vec{\theta}, z )}{\vec{g\,}^2}\bigg) \, \vec{g}.
\label{farhar23}
\end{equation}
Part of the art of successful machine learning is to judiciously set the learning rate which may  vary as the learning proceeds.

We do not yet have a quantum computer at our disposal but we can simulate the quantum process using a conventional computer.  Of course this is only possible at a small number of bits because the Hilbert space dimension is $2^{(n+1)}$.   The simulation has the nice feature that once the quantum state  \eqref{farhar4} is computed, we can evaluate the expected value of $Y_{n+1}$ directly without doing any measurements.  Also for the systems that we simulate, the individual unitaries are of the form \eqref{farhar7} and we can directly evaluate expression \eqref{farhar20c}. So in our simulations we evaluate the gradient exactly without recourse to finite difference methods.

\subsection{Learning Subset Parity}
As our first example we consider learning subset parity.  Recall that given a subset $\mathscr{S}$ the unitary $U_{P_{\mathscr{S}}}$ given by \eqref{farhar15} will express subset parity on all input strings with zero sample loss.  To learn we need a set of unitaries that depend on parameters with the property that for each subset $\mathscr{S}$ there is a parameter setting that produces $U_{P_{\mathscr{S}}}$.  A simple way to achieve this is to use $n$ parameters
\begin{equation}
U( \vec{\theta}\,) = \exp \Big( i \, \tfrac{\pi}{4} \, X_{n+1} \Big) \, \exp \Big(- i\, \sum\limits_j\, \theta_j  \, B_j  \, X_{n+1}\Big)
\label{farhar24}
\end{equation}
and we see that the representation is perfect with $\theta_j = \frac{\pi}{2}$ if $j$ is in the subset and $\theta_j = 0$ if $j$ is not in the subset.  We set up a numerical simulation to see if we could learn these optimal parameters.  Working from 6 to 16 bits, starting with a random $\vec{\theta}$ we found that with stochastic gradient descent we could learn the subset parity label function with far fewer than $2^n$ samples and therefore could successfully predict the label of unseen examples.  We also found that introducing 10\% label noise did not impede the learning.

        However what we just described was success at low bit number and we now argue that as the number of bits gets larger subset parity becomes impossible to learn.  To see this first note that we can explicitly compute the expected value of $Y_{n+1}$
\begin{equation}
\bra{z , 1} U^\dagger\, (\vec{\,\theta}\, ) Y_{n+1} U\, (\vec{\,\theta}\, ) \ket{z, 1}  =  \cos \Big(2 \sum\limits_j\, \theta_j \, b_j \Big)   .
\label{farhar25}
\end{equation}
With the label $l(z)$ we can plug this into the sample loss given by \eqref{farhar5}.  But now we can compute the average of the sample loss over all $2^n$ string since we have explicit formulas for the label and the expectation  \eqref{farhar25}.   There are different, but similar looking, formulas depending on the value of $n$mod4 and the number of bits in the set $\mathscr{S}$.  To be concrete consider the case that $n$ is a multiple of 4 and the set $\mathscr{S}$ contains all $n$ bits.  In this case the average over all inputs of the sample loss,  called the empirical risk, is
\begin{equation}
  1 - \cos (\theta_1 + \theta_2 + \ldots  \theta_n) \sin(\theta_1) \sin(\theta_2) \dots \sin (\theta_n).
  \label{farhar26}
\end{equation}
We see that this achieves its minimum when all $\theta$'s are $\frac{\pi}{2}$.  Imagine searching for the minimum of this function over say $[0\, \pi]^n$.  The function just displayed is exponentially close to 1 except in an exponentially small subvolume centered at the optimal angles.  Accordingly the gradient is exponentially small except near the optimal angles.  So even if we had access to the empirical risk no gradient based method could be used to find the optimal angles since for sufficiently large $n$ the gradients will fall beneath machine precision.  Of course for a given training example, the gradient of loss$(\vec{\theta} , z )$ will not typically be small. But the fact that the average is near zero means that there is not enough bias in the individual gradients for a stochastic gradient descent algorithm to drift to the tiny region near the optimal angles.

        What we have just illustrated is a specific example of a phenomena explored in the paper ``Failure of Gradient Based Deep Learning" \cite{Failures}.  Here the authors consider machine learning in the situation where the network can express (or represent if you prefer) a large set of label functions.  Different settings of the weights give rise to the different functions.  They have the restriction that the functions are orthonormal.  In this setting they show that the gradient of the empirical risk at (almost) any value of the weights is independent of which function is used to label the data.  This means that the gradient cannot be used to distinguish the label functions.  In our case we have $2^n$ different subset parity functions and they are indeed orthonormal so we fall under the curse of the paper and our approach is doomed at large bit number.  As long as we stick to our basic setup we cannot break the spell. 
        
        A word of caution about learning subset parity.  In the classical setting if you have $n$ linearly independent labeled data strings, then you can use linear algebra to identify the subset associated with the label. Once the subset is identified generalization to all other inputs is immediate.  In the quantum setting if the data is presented as a uniform superposition over all strings with the coefficient of each term having a factor of $+1$ or $-1$ given by the label, then the subset can be found after performing a Hadamard on the superposition state \cite{ParityIBM}. (Preparing this state takes two calls of the label function acting on the uniform superposition with all  $+1$ coefficients.) However we do not know how to apply this trick to other label functions.
\bigskip

\subsection{Learning Subset Majority}

Recall that we can represent subset majority with the unitary operator \eqref{farhar17} with $\beta$ set to  $0.9 \pi / n$.  By thresholding the expected value of $Y_{n+1}$, as per \eqref{farhar20}, we can achieve zero categorical error.  We are interested in having a parameter dependent unitary for which there are parameter settings corresponding to the different subsets. Consider the unitary
\begin{equation}
U(\vec{\theta})= \exp\Big({i \frac{\beta}{2}  \sum_j \theta_j Z_j X_{n+1}}\Big)
\end{equation}
Now with $\theta_j = 1$ if $j$ is in the subset and $\theta_j = 0$ if $j$ is not in the subset then $U(\vec{\theta})$ represents subset majority on the selected subset.  We now ask if we can learn the correct $\theta$'s given a training set labeled according to the majority of a selected subset. Note that the predicted label value on a training sample $z$ is 
\begin{equation}
\sin(\beta  \sum_j \theta_j  z_j)
\end{equation}
and so rounding up or down gives the predicted label as
\begin{equation}
\text{sign} (\sum_j \theta_j z_j) .
\end{equation}
This result has a direct interpretation in terms of a classical neural network with a single neuron that depends on the weights $\vec{\theta}$.  The $\sum_j \theta_j z_j$ is the result of the neuron acting on the input.  The nonlinearity comes from applying the sign function.  But we know that since the data can be separated with one hyperplane, this is an easy label to learn with a single neuron. The same reasoning leads us to the conclusion that our QNN can efficiently be trained to represent subset majority. We indeed saw in small scale numerical simulations that the QNN was able to learn subset majority with low sample complexity.
\bigskip

\subsection{Learning to Distinguish Digits}
Classical neural networks can classify hand written digits.  The archetypical example comes from the MNIST data set which consists of 55,000 training samples that are 28 by 28 pixilated images of hand written digits that have been labeled by humans as representing one of the ten digits from 0 to 9 \cite{MNIST}.  Many introductory classes in machine learning use this data set as a testbed for studying simple neural networks.  So it seems natural for us to see if our quantum neural network can handle the MNIST data.  There is no obvious way to attack this analytically so we resort to simulation.  The limitation here is that we can only easily handle say 16 bit data using a classical simulator of a 17 qubit quantum computer with one readout bit.   So we use a downsampled version of the MNIST data which consists of 4 by 4 pixilated images.  With one readout bit we cannot label ten digits so instead we pick two digits, say 3 and 6, and reduce the data set to consist of only those samples labeled as 3 or 6 and ask if the quantum network can distinguish the input samples.

The 55,000 training samples break into groups of roughly 5,500 samples for each digit.  But upon closer examination we see that the samples corresponding to say the digit 3, consist of  797 distinct 16 bit strings while for the digit 6 there are 617 distinct 16 bit strings.  The images are blurry and in fact there are 197 distinct strings that are labeled as both 3 and 6.  For our digit distinction task we decided to reduce the Bayes error to 0 by removing the ambiguous strings.  Going back to the 5,500 samples for each digit and removing ambiguous strings, leaves 3514 samples that are labeled as 3's and 2517 that are labeled as 6's. We combine these to make a training set of 6031 samples

As a preliminary step we present the labeled samples to a classical neural network.  Here we run a Matlab classifier with one internal layer consisting of 10 neurons.  Each neuron has 16 coefficient weights and one bias weight so there are 170 parameters on the internal layer and 4 on the output layer.  The classical network has no trouble finding weights that give less than one percent classification error on the training set.  The Matlab program also looks at the generalization error but to do so it picks a random 15 percent of the input data to use for a test set. Since the input data set has repeated occurrences of the same 16 bit strings, the test set is not purely unseen examples.  Still the generalization error is less than one percent.

We now turn to the quantum classifier.  Here we have little guidance as to how to design the quantum circuit.  We decided to restrict our toolkit of unitaries to consist of one and two qubit operators of the form (7).  We take the one qubit $\Sigma$'s to be $X, Y$ and $Z$ acting on any of the 17 qubits.  For the two qubit unitaries we take $\Sigma$ to be $XY, YZ, ZX, XX, YY$ and $ZZ$ between any pair of different qubits. The first thing we tried was a random selection of 500 (or 1000) of these unitaries. The randomness pertains to which of the 9 gate types are picked as well as to which qubits the gates are applied to.  Starting with a random set of 500  (or 1000) angles, after presenting a few hundred training samples, the categorical error settled in at around 10 percent.  But the sample loss for individual strings was typically only a bit below 1 which corresponds to a quantum success probability of just over 50 percent for most stings.  Here the trend was in the right direction but we were hoping to do better.

After some playing around we tried restricting our gate set to $ZX$ and $XX$ with the second qubit always being the readout qubit and the first qubit being one of the other 16. The motivation here is that the associated unitaries effectively rotate the readout qubit around the $x$ direction by an amount controlled by the data qubits.  A full layer of $ZX$ has 16 parameters as does a full layer of $XX$.  We tried an alternation of 3 layers of $ZX$ with 3 layers of $XX$ for a total of 96 parameters.  Here we found that starting from a random set of angles we could achieve two percent categorical error after seeing less than the full sample set.

The accomplishment here is that we demonstrated that a quantum neural network could learn to classify real world data.  Admittedly the data set could easily be classified by a classical network.  And working at a fixed low number of bits precludes any discussion of scaling.  But our work is exploratory and without much effort we have a quantum circuit that can classify real world data.  Now the task is to refine the quantum neural network so it performs better. Hopefully we can find some principles (or just inspiration) that guides the choice of gate sets.

\subsection{Classical Data Presented in Superposition}
We have focused on supervised learning of classical data where the data is presented to the quantum neural network one sample string at a time.  With quantum resources it seems natural to ask if the data can be presented as quantum states that are superpositions of computational basis states that are associated with batches of samples \cite{PhysRevLett.113.130503}.  Again focus on binary classification.  We can divide the sample space into those samples labeled as $+1$ and those labeled as $-1$.  Consider the states
\begin{equation}
\ket{+1} = N_+ \sum\limits_{z: l (z) = 1} e^{i\varphi_z} \ket{z,1} \hspace{2em} \ket{-1} = N_- \!\!\sum\limits_{z: l (z) = -1} e^{i\varphi_z} \ket{z,1}
\label{farhar34}
\end{equation}
where $N_+$ and $N_-$ are normalization factors.  At this point we have no inspired choice for the phases and in our example below we just set them all to 0.  Each of these states can be viewed as a batch containing all of the samples with the same label. Note that QNNs naturally offer two distinct strategies for batch learning. In \eqref{farhar34}
we combine different samples into single superposition states and then evaluate the gradient on a suitable loss function. Alternatively we can compute the gradient of \eqref{farhar5} one sample at a time and then average the gradients as done in traditional batch learning. Here we concentrate on the first approach.

Return to equation \eqref{farhar8} which gives the unitary associated with any label function.  Note that the expected value of this operator of the state  $\ket{+1}$  is $+1$ whereas the expected value of the state $\ket{-1}$ is $-1$.  This is because the unitary is diagonal in the computational basis of the data qubits so the cross terms vanish and the phases are irrelevant.   Now consider a parameter dependent unitary $U(\vec{\theta})$ which is diagonal in the computational basis of the data qubits.  The expected value of $Y_{n+1}$ of the state obtained by having this  operator act on $\ket{+1}$  is the average over all samples with the label  $+1$ of the quantum neural network's predicted label values.  Similarly for the state $\ket{-1}$.   In other words if $U(\vec{\theta})$ is diagonal in the computational basis of the data qubits then
\begin{equation}
1-{\frac{1}{2}}\Big(\bra{+1} U^\dagger (\vec{\theta}) Y_{n+1} U (\vec{\theta}) \ket{+1}  - \bra{-1} U^\dagger (\vec{\theta}) Y_{n+1} U (\vec{\theta})\ket{-1}\Big)
\label{farhar35}
\end{equation}
is the empirical risk of the whole sample space.  If parameters $\vec{\theta}$ are found that make this 0, then the quantum neural network will correctly predict the label of any input from the training set. 

To test this idea we looked at the two digit distinction task described in the previous section.  The state $\ket{ +1}$ was formed by superimposing computational basis states corresponding to all the strings labeled as the digit 3.  We chose the phases to all be 0.  Note that strings recur so that different basis states were added in with different weights.  Similarly the state $\ket{-1}$  was the quantum superposition of the basis states corresponding to strings labeled as 6.  To start we used a gate set that was diagonal in the computational basis of the data qubits.  A simple choice was $ZX$ and $ZZX$ with the $Z$�s always acting on data qubits and the X acting on the readout.  This gave rise to $16 + 16 \cdot15/2 = 136$ parameters. With these gates expression \eqref{farhar35} is the empirical risk of the full data set for this quantum neural network.   For a given choice of input parameters we could numerically evaluate \eqref{farhar35}.  Starting from a random choice of parameters we performed gradient descent to decrease the empirical risk.  The empirical risk settled down to a value of around .5.  Recall that with our conventions a value of 1 corresponds to random guessing and 0 is perfection.  We tested the learning by looking at the categorical error on a random sample of input strings.  Here the error was a few percent.  Note that we did not follow good machine learning methodology since our test strings were included in the superposition states.  But the point here is that we could set up a quantum neural network that can be trained with superposition states of real world data.

We also expanded the gate set beyond those that are diagonal in the data qubit computational basis.  Now expression \eqref{farhar35}  can no longer be directly read as the empirical risk of the quantum neural network acting on the whole sample space.  Still driving it to a low value at least means that the states $\ket{+1}$ and $\ket{-1}$ are correctly labelled.  We used the gate set that we used in classifying the MNIST data one sample at a time. These are $XX$ and $ZX$ gates where the first qubit is a data qubit and the second  always acts on the readout. We were again able to drive \eqref{farhar35} to a value of around .5 and  the quantum neural network had low categorical error on a test set of individual data samples. 

It is natural to ask if learning with quantum batches, that is, superpositions of computational basis states corresponding to strings with the same label is “better’ than learning by presenting sequentially states corresponding to single labeled strings.  If we track the individual sample loss as new training examples are presented in the non-superposition case we see the sample loss fluctuate seemingly randomly until it trends down on average to a low value. In the quantum batch case, if we follow the progress of the empirical risk, it smoothly decreases until it settles at a local minimum. We did numerical experiments to contrast the quantum batch learning with individual sample learning. With 16 bit data working on a 17 qubit simulator, we saw more than an order of magnitude improvement in the sample complexity required to get comparable (or better) generalization error on individual test samples. Given that the empirical risk \eqref{farhar35}  in the quantum batch case is a “smoother” function of parameters than the individual sample loss \eqref{farhar5} there may be better strategies to minimize it than the gradient descent method we adopted. 

\subsection{Learning a Property of Quantum States}
So far we have focused on using a quantum neural network to learn labels from classical data. The sample data is encoded in a quantum state, either a computational basis state associated with a data string or a superposition of such states.  But with a quantum network, it is natural to input  general quantum states  with the hope of learning to classify a label that is a property of the quantum state.  Now there is no classical neural network that can attempt the task since classical networks do not accept quantum states as inputs.   The basic idea is to present an $n$-qubit state $\ket{\psi}$  to the quantum network with the readout qubit set to 1 as before.  So given a unitary $U (\vec{\theta})$, we make the state
\begin{equation}
U(\vec{\theta}) \ket{\psi , 1}
\label{farhar27}
\end{equation}
and then measure $Y_{n+1}$. The goal is to make the outcome of the measurement correspond to some two valued label of the state.  We now turn to an example.

Consider a Hamiltonian $H$,  that is a sum of local terms with the additional assumption that it is traceless so we know that there are positive and negative eigenvalues.  Given any quantum state $\ket{\psi}$ we label the state according to whether the expected value of the Hamiltonian is positive or negative:
\begin{equation}
l \big( \ket{\psi} \big) = \text{sign} \, \big( \brakets{\psi}{H}{\psi} \big) .
\label{farhar28}
\end{equation}
Consider the operator
\begin{equation}
U_H ( \beta) = \exp ( i \beta\, H\,  X_{n+1} )
\label{farhar29}
\end{equation}
where we take $\beta$ to be small and positive.  Now
\begin{equation}
\bra{\psi ,1}U^\dagger_H (\beta)  Y_{n+1} U_H (\beta) \ket{\psi, 1}  = \brakets{\psi}{\sin ( 2 \beta  H )}{\psi}
\label{farhar30}
\end{equation}
so for sufficiently small $\beta$ this is close to
\begin{equation}
2 \beta \brakets{\psi}{H}{\psi}
\label{farhar31}
\end{equation}
and we have the sign of the expected value of our predicted label agreeing with the true label.  In this sense we have expressed the label function with a quantum circuit with small categorical error. The error arises because the right hand side of \eqref{farhar30} is only approximately equal to \eqref{farhar31}. However if we take $\beta$ to be much less than the inverse of the norm of $H$, we can make the error small.

  To be concrete consider a graph where on each edge  we have a $ZZ$ coupling with a coefficient of +1 or -1 randomly chosen. The Hamiltonian is
\begin{equation}
H = \sum\limits_{<i j>}  J_{ij}\, Z_i\, Z_j
\label{farhar32}
\end{equation}
where the sum is restricted to edges in the graph and $J_{ij}$ is +1 or -1.    Suppose there are $M$ terms in $H$.  We can first pick $M$ angles $\theta_{ij}$ and consider circuits that implement unitaries of the form:
\begin{equation}
U ( \vec{\theta} ) = \exp\, \Big( i \sum\limits_{<i j>} \theta_{ij}\, Z_i\, Z_j\, X_{n+1} \Big)
\label{farhar33}
\end{equation}
If we pick $\theta_{ij} = \beta J_{ij}$  we have the operator $U_H (\beta)$ given in \eqref{farhar29} which ensures that we can represent the label \eqref{farhar28} by picking $\beta$ small.  We can ask if we can learn these weights.

Our quantum states $\ket{\psi}$ live in a $2^n$ dimensional Hilbert space and we may not expect to be able to learn to correctly label all of these states.  The Hamiltonian we consider has bit structure so we might restrict to quantum states that also have bit structure. For example, they could be built by applying few qubit unitaries to some simple product state. In this case we would only present training states of this form and only test our circuit on states of this form.

We did a simple numerical experiment that we report here.  Working with 8 data qubits and one output qubit we tossed a random 3 regular graph which accordingly has 12 edges.  In this case there were 12 parameters $\theta_{ij}$ used to form the operator \eqref{farhar33}.   For our training states we used product states that depend on 8 random angles. The state is formed by rotating each of the 8 qubits, which each start as an eigenstate of the associated $X$ operator, about the $y$ axis by the associated random angle.  Test states are formed in the same manner.  Since the states are chosen randomly from a continuum we can be confident that the training set and test set are distinct.  After presenting roughly 1000 test states the quantum network correctly labels 97\% of the test states.   We expanded the class of unitaries to include more parameters.  Here we introduced two layers of $XX$ and $ZX$ unitaries of the form (7) where the first operator acts on one of the 8 data qubits and the second operator acts on the readout qubit.  This introduced another 32 parameters for a total of 44 parameters.  We found that our learning procedure again could achieve 3\% categorical error after seeing roughly 1000 training examples.

Our simple numerical example demonstrates that it is possible for a quantum neural network to learn to classify labeled quantum states that come from a subset of the Hilbert space.  The learning generalizes beyond the training set. A cautionary note: The states that we use to train and test the QNN come from simple unitaries acting on a simple product state.  Therefore there is a classical description of these states.  One can imagine a classical competitor to the QNN that takes a description of the input state and a description of the the quantum Hamiltonian and asks if the label (42) can be learned by a purely classical process.  But hopefully what we have demonstrated here will find use in quantum metrology or in other uses of quantum networks classifying quantum states for which a compact classical description of the quantum state is not available.

\section{Conclusions and Outlook}
In the near future we will have gate model quantum computers with a sufficient number of qubits and sufficiently high gate fidelity to run circuits with enough depth to perform tasks that cannot be simulated on classical computers \cite{Otterbach}\cite{NewIBMHardware}\cite{2}\cite{3}\cite{4}.  One approach to designing quantum algorithms to run on such devices is to let the architecture of the hardware determine which gate sets to use \cite{2017arXiv170306199F}. In this paper, in contrast to prior work \cite{QuantumNeuron}\cite{QNNOxford}, we set up a general framework for supervised learning on quantum devices that is particularly well suited for implementation on quantum processors what we hope to have available in the near term. 

To start we showed a general framework for using a quantum device to classify classical data.  With labeled classical data as inputs, we map an input string to a computational basis state that is presented to the quantum device.  A Pauli operator is then measured on a readout qubit. The goal is to make the measurement outcome correspond to the correct binary label of the input string.  We showed how to design a quantum neural network that can in principle represent all Boolean label functions of $n$ bit data.  The circuit is a sequence of two qubit unitaries but for certain label functions, the circuit may need to be exponentially long in $n$.  This representation result, analogous to the classical representation result, allows us to turn to the question of learning.

The quantum circuits that we imagine running are sequences of say one and two qubit unitaries each of which depends on a few parameters.  These unitaries may come from the toolkit of gates provided by the experimentalist. Without error correction, the number of gates that can be applied is limited by the individual gate fidelities and the final error that can be tolerated.  In a theoretical investigation we can imagine having access to any set of local gates all of which work perfectly.  Still with a given set of gates it is not clear which label functions can be represented or learned.  To learn we start with a parameterized gate set and present data samples to the quantum device.  The output qubit is measured, perhaps repeatedly for good accuracy.  The parameters are then varied slightly to make it more likely that the output corresponds to the correct label. This method of updating parameters (or weights in machine language parlance) is standard practice in leaning to classify data in the traditional supervised learning setting. The ultimate goal is to be able to correctly classify data that has not been seen during training. 

As a demonstration of our approach we looked at using a quantum neural network to distinguish digits.  Here we used a data set of downsampled images of two different handwritten digits.  Each image consists of 16 data bits and a label saying which of the two digits the image represents.  Without access yet to an actual 17 qubit error free quantum computer, we ran a classical computer as a simulator of a quantum device.  We picked a simple parameterized gate set and started with random parameters.  Using stochastic gradient descent we were able to learn parameters that labeled the data with small error.  This exercise served as a proof of principle that our quantum methodology could be used to classify real world data.

With a quantum network it seems natural to attempt to present classical data in superposition.  A single quantum state that is a superposition of computational basis states each of which represents a single sample from a ``batch" of samples, can be viewed as quantum encoding of the batch.  Here different phases on the components give rise to different quantum states.  Returning to our digit distinction example, we formed two quantum states, each of which is a uniform (zero phase angle) superposition of all the data samples corresponding to one of the selected digits.  Either state can be presented to the quantum neural network.  We can then measure the difference in the expected value of $Y_{n+1}$ between the two states. This is the quantum analog of the empirical risk.  With the simulator at hand we numerically evaluate this difference which is a one shot picture of all of the input samples.  Doing gradient descent on the parameters we found parameter values that give a quantum circuit with low categorical error on test data.

Of course classical data does not come in quantum superpositions.  So to run the previous protocol, the data must be read into a quantum device that prepares the superposition.  The superposition is consumed by the quantum network so to run the gradient descent fresh copies of the superposition must be made.  Or perhaps many copies of the superposition can be made at the outset and stored for later use. Here is room to explore strategies that reduce computational cost.

We also looked at learning a label that is an attribute of a general quantum state.  Here the idea is to present quantum states to the quantum neural network and train the network to predict the label of an unseen example state.  To be concrete we considered a quantum system with a local Hamiltonian with positive and negative eigenvalues. The label of a quantum state is $+1$ if the expected value of the Hamiltonian is positive and the label is $-1$ if the expected value if negative.  All states in the $2^n$ dimensional Hilbert space have such a label.  But we restrict our attention to those states that can be made by applying local unitary transformations to a simple product state such as the all 0 computational basis state. In this way the states being explored have the same bit structure as the Hamiltonian.  We did a numerical simulation at $8+1$ qubits to see if this idea could get off the ground.  Here we presented product states to parameterized circuits.  We found that parameters could be learned that allowed the quantum network to correctly predict the sign of the expected value of the Hamiltonian on new examples.

Our work sets out a specific framework for building quantum neural networks that can be used to do supervised learning both on classical and quantum data.  Our numerical work was limited by the fact that only small quantum devices can be simulated on classical machines.  With greater resources and effort than we were willing to expend, one could go to say 40 qubit simulations.  But our simulations were exploratory and with little (or no) guidance as how to pick gate sets we adopted a strategy of limited simulation time per trial so we could try many different circuits.  Adopting this approach, we worked at no more than 17 qubits.  Based on our numerics we cannot make a case for any quantum advantage over classical competitors for supervised learning. Of course with labeled quantum states as input, complex enough that no concise classical description exists, there is no classical counterpart so the comparison cannot be made.  What we have done is to demonstrate that quantum neural networks can be used to classify data.

Still our framework will hopefully be implemented on a near term quantum device that cannot be classically simulated. There may be tasks that only the quantum device can handle. At big enough sizes it may be possible to see an advantage of the quantum approach over corresponding classical approaches. Quantum processors available in the near term may only allow for a modest number of input variables. In this case the application of a QNN to real world tasks can be hastened by using a classical-quantum hybrid architecture in which the first layers of the neural net are implemented classically and only the final layers which are often smaller are implemented as a QNN. 

Traditional machine learning took many years from its inception until a general framework for supervised learning was established.  We are at the exploratory stage in the design of quantum neural networks. Refer back to Fig.1. In our framework a quantum state serves as input and a single qubit is measured to give the predicted label. Note that we do not make use of the information contained in qubits $1$ to $n$. Doing so could inspire novel network designs \cite{2017QS&T....2d5001R}. There are endless choices for the quantum circuit that sits between the input and the output. We hope that others will discover choices that lead to establishing the full power of quantum neural networks. 

\section*{Acknowledgments}

We would like to thank Larry Abbott, Jeffrey Goldstone and Samuel Gutmann for sharing their human intelligence. We want to thank Matthew Coudron for valuable discussions during his summer internship at Google, Vincent Vanhoucke for encouraging us to apply the QNN to the low resolution version of MNIST, Vasil Denchev for implementation assistance and Nan Ding for sharing his machine learning wisdom. We also thank Charles Suggs for technical assistance. EF was partially supported by the National Science Foundation under grant contract number CCF-1525130 and by the Army Research Office under contract W911NF-17-1-0433.

\nocite{*} 
\printbibliography

\end{document}